# Diffusion-limited aggregation: A relationship between surface thermodynamics and crystal morphology

Vladislav A. Bogoyavlenskiy* and Natasha A. Chernova
*Low Temperature Physics Department, Moscow State University, Moscow 119899, Russia*
(Received 23 June 1999)

We have combined the original diffusion-limited aggregation model introduced by Witten and Sander with the surface thermodynamics of the growing solid aggregate. The theory is based on the consideration of the surface chemical potential as a thermodynamic function of the temperature and nearest-neighbor configuration. The Monte Carlo simulations on a two-dimensional square lattice produce the broad range of shapes such as fractal dendritic structures, densely branching patterns, and compact aggregates. The morphology diagram illustrating the relationship between the model parameters and cluster geometry is presented and discussed.

PACS number(s): 81.10.Aj, 02.70.Lq, 68.70.+w, 82.65.Dp

## I. INTRODUCTION

Diffusion-limited growth processes occur in a broad range of interesting systems ranging from physics to chemistry, material science, and biology [1–6]. The common features that we observe in these systems are the formation of complex interfacial structures that vary from compact to fractal [7,8]. The growth processes are described by nonlinear partial-differential equations and both the analytical and the numerical treatments of these equations are extremely difficult even on current computers [9,10]. As a result, many of the questions concerning structure formation and transitions between different growth morphologies have not so far been satisfactorily answered. Much effort has especially been devoted to establishing the relationship between cluster morphology and the growth mechanism.

The diffusion-limited aggregation (DLA) model introduced by Witten and Sander in 1981 [11] has attracted much attention because of the variety of growth shapes that it can produce [12,13]. The standard DLA model simulates the growth of an aggregate by considering the random walk of a particle on a lattice containing a seed. If the mobile particle encounters the seed, it ceases to move. As successive walkers repeat this process, the fractal aggregates are produced. In order to study more realistic DLA-processes, various kinds of aggregation models have been introduced. The most known of them consider sticking probability kinetics [14–18], surface diffusion [19–22,27], and many-particle interactions [23–27].

In this work, the modified DLA model based on surface thermodynamics of solid aggregate growing from vapor phase is investigated. The theory considers the surface chemical potential as a thermodynamic function of temperature and nearest-neighbor configuration. Monte Carlo (MC) simulations on a two-dimensional (2D) square lattice produce the variety of growth patterns such as fractal dendritic structures, densely branching patterns, and compact aggregates, i.e., three main morphological types observed in crystal growth. The paper is organized as follows. In Sec. II the general model of surface thermodynamics and mass transfer is formulated. The subject of Sec. III is numerical procedure and results of MC simulations. Finally, in Sec. IV we discuss the observed growth patterns in terms of a proposed morphological diagram.

## II. GENERAL MODEL

Let us assume that the motion and aggregation of the growth units take place on a square 2D grid, and restrict our study to a physical system with the following properties. (i) The nutrient vapor phase consists of two components: the growth species and an inert gas which randomizes the motion of the growth units. (ii) The growth unit is transported towards the surface of the solid aggregate only by diffusion; there is no convective motion of the nutrient. (iii) The heat transfer realizes through the solid phase from the surface to the cool origin. (iv) The probability $p_{\text{growth}}$ that the growth unit sticks onto a vacant surface position is given by the thermodynamic condition

$$p_{\text{growth}} = 1 \quad \text{if} \quad \Delta\mu < 0, \quad (1)$$

$$p_{\text{growth}} = 0 \quad \text{if} \quad \Delta\mu \geq 0, \quad (2)$$

where $\Delta\mu \equiv \mu_{\text{solid}} - \mu_{\text{vapor}}$ is the chemical potential difference between solid and vapor states.

The thermodynamics of the aggregation is considered to be the following: (i) the vapor chemical potential $\mu_{\text{vapor}}$ is constant and (ii) the chemical potential of the solid phase is the thermodynamic function of the temperature $T$ and local surface configuration $\Sigma$,

$$\Delta\mu(T,\Sigma) = \frac{L}{T_A}(T - T_A) + \Sigma. \quad (3)$$

Here $L=$const is the latent heat and $T_A$ is the equilibrium temperature of the aggregation. Assuming only nearest-neighbor interactions on the square lattice, the linear form of the local surface configuration $\Sigma$ can be represented as

$$\Sigma(n) = \frac{L}{T_A}(2-n)T_{\text{surf}}, \quad (4)$$

---

*Electronic address: bogoyavlenskiy@usa.net

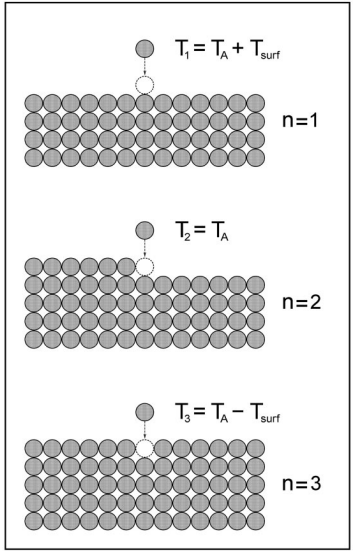

FIG. 1. Three equilibrium temperatures of aggregation corresponding to three local configurations of vacant surface position on a square lattice.

where $n$ is the number of occupied nearest neighbors of the vacant surface position and $T_{\text{surf}}=\text{const}$ is the configuration increment. The physics of the relations (3) and (4) becomes clear if we rewrite the equilibrium condition of the aggregation $\Delta\mu=0$ as

$$T_n = T_A + (n-2)T_{\text{surf}}. \quad (5)$$

Here $T_n$, $n=1\cdots 3$ are the equilibrium temperatures of the aggregation onto one-, two-, and three-neighboring configurations of the vacant surface position, respectively. The overall picture of the surface thermodynamics is summarized by Fig. 1 which shows the aggregation of the growth units on the square lattice.

The surface aggregation releases the heat which diffuses to the cool origin. The heat transfer obeys the diffusion equation and appropriate boundary condition

$$\frac{\partial T(\mathbf{r},t)}{\partial t} = D\nabla^2 T(\mathbf{r},t) + \frac{L}{C}\frac{\partial N_A(\mathbf{r},t)}{\partial t}, \quad (6)$$

$$T(0,t) = T_O. \quad (7)$$

Here $\mathbf{r}=(x,y)$ is the coordinate vector, $D=\text{const}$ is the thermal diffusivity of the solid phase (we consider $D=0$ in the vapor phase), $C=\text{const}$ is the specific heat, $N_A(\mathbf{r},t)$ is the density of the aggregating units, and $T_O=\text{const}$ is the origin temperature.

The set of Eqs. (3)–(7) completely describes the surface thermodynamics and the heat transfer. For MC simulations, we define the additional parameters

$$\lambda \equiv \frac{L}{C(T_A-T_O)}, \quad (8)$$

$$\xi \equiv \frac{T_{\text{surf}}}{(T_A-T_O)}, \quad (9)$$

where $\lambda$ and $\xi$ are the dimensionless latent heat and surface energy, respectively. To describe the mass transfer, let us assume that the number of growth units reaching the aggregate surface per time unit (i.e., the intensity of the mass transfer) $N_{\text{growth}}$ is constant. Then we can introduce the dimensionless diffusivity as

$$\eta \equiv \frac{D}{a^2 N_{\text{growth}}}, \quad (10)$$

where $a$ is the parameter of the square lattice. It is clear that diffusivity $\eta$ establishes the ratio of the heat and mass transfer.

## III. NUMERICAL SIMULATIONS

According to our assumption, the motion of growth units is governed only by diffusion. The diffusive motion is described by a simple random walk with the isotropic jump length $a$. This condition on the square lattice is given by the formula

$$P_{k+1}(x,y) = \frac{1}{4}\{P_k(x+a,y) + P_k(x-a,y) + P_k(x,y+a) + P_k(x,y-a)\}, \quad (11)$$

where $P_k(x,y)$ is the probability that a growth unit can be found at location $(x,y)$ after $k$ steps of its motion.

The problem is consistent with the following MC simulation on the $200a \times 200a$ square lattice. Initially, a nucleus is located at the origin. A random walker is released from a circular source outside the central cluster. The source location is considered to be $R_G+5a$ where $R_G$ is the radius of gyration of the growing cluster. This restriction of random walks saves considerable computer time [28] so this simplification is used. The random walk is pursued until the walker encounters a vacant site on the aggregate surface. After that the nearest-neighboring configuration and the temperature of the site are checked. According to Eqs. (3), (4), the value of $\Delta\mu$ is calculated. If $\Delta\mu<0$, a registration of the site is made. In the opposite case ($\Delta\mu \geq 0$) that walker is disregarded. When a site has been registered 200 times, it is considered to be occupied. This number of registrations has the effect of reducing the noise inherent in simulations of this kind [29–31]. When the walker aggregates onto the surface site, its temperature is initially considered to be $T_A+L/C$, and then decreases according to the heat transfer equations (6),(7). The simulation is continued until the aggregate reaches a specified size ($R_G=100a$).

The results of MC simulations are presented in Figs. 2 and 3. Figure 2 shows the morphological evolution of the growing aggregate caused by the variation of the diffusivity $\eta$ (the parameters $\lambda$ and $\xi$ are considered to be constant). At low values of $\eta$ the compact structure is observed [Fig. 2(a)]. The compact growth is characterized by epitaxial aggregation of the growth units so the shape is square. The increase of $\eta$ results in the morphological transition from compact to densely branching morphology (DBM). The DBM phase locally resembles the ramified structure of a DLA fractal but at larger length scale it is densely packed and the pattern has a well defined smooth envelope. At the beginning of the transition the pattern looks similar to a square cracked along its symmetry axis [Fig. 2(b)]. Then the DBM aggregate becomes more isotropic [Figs. 2(c) and 2(d)]. When the value of $\eta$ exceeds the second critical point the DBM structure transforms to the fractal [Figs. 2(e) and 2(f)]. This transition is connected to the change of the envelope shape from convex to concave. The aggregate becomes tip-stable dendritic with the fourfold symmetry [Fig. 2(f)]. This structure is quite similar to one obtained by the standard noise-reduced DLA algorithm [13,18,30]. The possible morphologies at the variation of the surface energy $\xi$ are summarized by Fig. 3. The fractal structures grow at low values of $\xi$ [Figs. 3(a) and 3(b)]. The increase of $\xi$ leads to the successive transitions fractal-DBM [Figs. 3(c) and 3(d)] and DBM-compact [Figs. 3(e) and 3(f)].

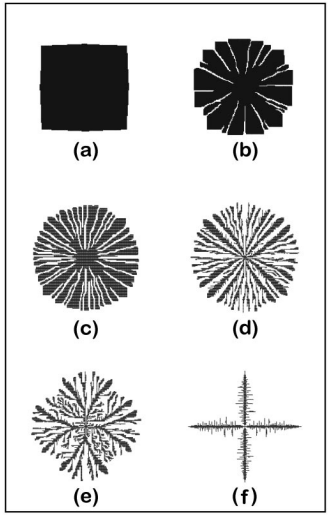

FIG. 2. Morphology of growing aggregate as a function of diffusivity $\eta$. These are results for $\lambda=0.3$ and $\xi=0.3$; values of $\eta$ are 0.02 (a), 0.04 (b), 0.1 (c), 0.4 (d), 1.0 (e), and 4.0 (f).

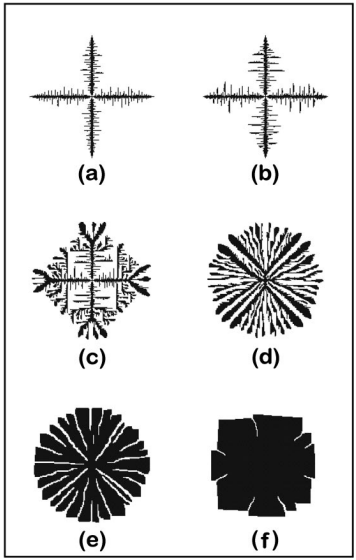

FIG. 3. Morphology of growing aggregate as a function of surface energy $\xi$. These are results for $\lambda=0.3$ and $\eta=2.0$; values of $\xi$ are 0.1 (a), 0.3 (b), 0.4 (c), 0.6 (d), 0.8 (e), and 0.9 (f).

## IV. DISCUSSION

The numerical simulations presented in the previous section demonstrate the broad range of the growth patterns. We have obtained three possible morphologies: compact, DBM, and fractal. The observed morphological transitions (fractal-DBM and DBM-compact) can be described by a morphology diagram in coordinates $(\xi,\lambda,\eta)$ where the phase fields of this diagram are defined as

$$\eta\{\text{fractal}\leftrightarrow\text{DBM}\}\equiv\eta_1(\xi,\lambda), \quad (12)$$

$$\eta\{\text{DBM}\leftrightarrow\text{compact}\}\equiv\eta_2(\xi,\lambda). \quad (13)$$

### A. Fractal-DBM transition

To determine the function $\eta_1(\xi,\lambda)$, let us find the solution of the heat transfer equation (6) in the case of the fractal dendritic growth. Because of high values of the parameter $\eta$, we can assume the quasistationary limit $\partial T/\partial t\to 0$ in the solid phase. The growth units aggregating onto the surface sites give only a slight temperature perturbation which quickly slows down. As a consequence, the temperature field in the solid phase can be written as

$$T(x)=T_O+(T_l-T_O)\frac{|x|}{l}, \quad (14)$$

where $l\gg a$ is the distance between a growth edge and the origin and $T_l\lesssim T_A-T_{\text{surf}}$ is the temperature of the growth edge (the inequality is a representation of a free aggregation condition onto all vacant surface positions; it is a necessary condition of the fractal growth). As a result, the cooling velocity of an aggregating growth unit $\partial T_{\text{unit}}/\partial t$ follows from the equation

$$\frac{\partial T_{\text{unit}}}{\partial t}=\frac{D}{a^2}(T_l-T_O)=\text{const}. \quad (15)$$

The initial temperature of the growth unit is $T_A+L/C$, the final temperature equals to the cool origin temperature $T_O$. So the time of this cooling $\Delta t_C$ is given by the relation

$$\Delta t_C=\frac{a^2(T_A+L/C-T_O)}{D(T_l-T_O)}. \quad (16)$$

Assuming the growth edge temperature is equal to the equilibrium of one corresponding to the aggregation of one-neighboring vacant position (i.e., $T_l=T_1=T_A-T_{\text{surf}}$), and then substituting Eq. (16) into Eq. (10), we obtain the following formula for the fractal-DBM transition:

$$\eta_1(\xi,\lambda)=\frac{T_A+L/C-T_O}{T_A-T_{\text{surf}}-T_O}=\frac{1+\lambda}{1-\xi}. \quad (17)$$

To illustrate the validity of our assumption (14), we investigated the real temperature fields during MC simulation. The numerical results of temperature fields in a fractal dendritic cluster are presented in Fig. 4. The figure shows that Eq. (14) is acceptable quite in all range of coordinate $x$, and only close to the growth edges the linear dependence trans-forms to a nonlinear one. The temperature perturbations caused by aggregating units are negligible.

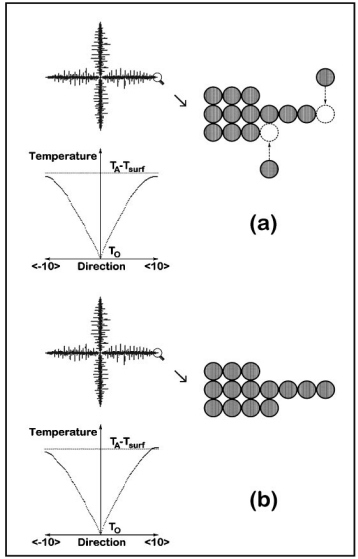

FIG. 4. Temperature fields in growing fractal dendritic aggregate before (a) and shortly after aggregation of growth units (b).

### B. Compact-DBM transition

The case of the compact-DBM transition is more complicated than the previous one because of nonlinear temperature oscillations on the surface. The overall picture of the compact growth [Fig. 2(a)] is summarized by Fig. 5 which shows the temperature fields in the solid phase at various growth stages. To show the temperature sequence of the pattern, we subdivided the solid phase into two fields: ''cool'' (light-gray color) and ''hot'' (dark-gray color). The temperature which differentiates the fields corresponds to the equilibrium of one for the aggregation onto one-neighboring vacant position:

$$T_1 = T_A - T_{\text{surf}}. \qquad (18)$$

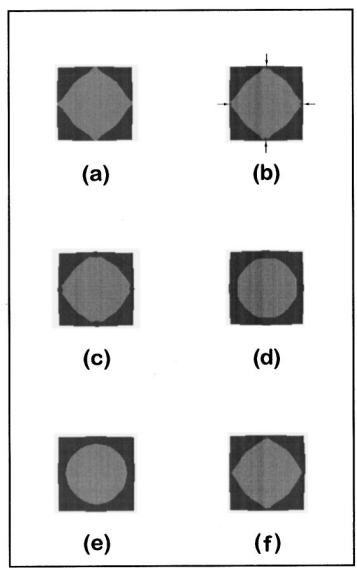

FIG. 5. Temperature fields in growing compact aggregate at different growth stages: before nucleation (a), nucleation (b), (c), epitaxial growth (d), (f). Sites with temperatures $T<T_1$ and $T>T_1$ are colored in light-gray and dark-gray, respectively. Arrows mark nucleation of monoparticle layer.

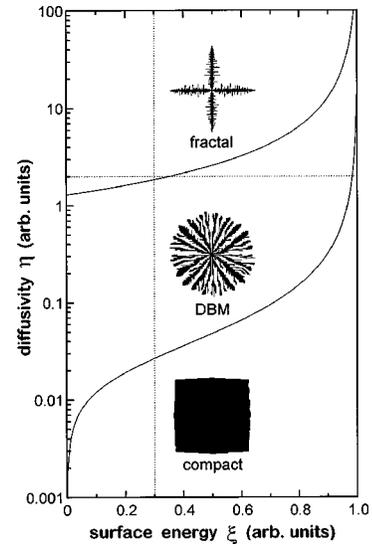

FIG. 6. Morphology diagram in coordinates $(\xi,\log\eta)$ obtained for $\lambda=0.3$. Dotted lines correspond to sections $\xi=0.3$ and $\eta=2.0$.

One can see that only ''cool'' surface sites are responsible for nucleation of new monoparticle layer. To illustrate this, let us discuss the main growth stages in details. Before the nucleation of a new monoparticle layer, the ''cool'' sites appear at the center of each crystal side [Fig. 5(a)]. Shortly after the nucleation the ''cool'' surface sites transform to ''hot'' ones due to the latent heat of the aggregation [Figs. 5(b) and 5(c)]. Then the monoparticle layer begin to grow from the center of the crystal side to the edge, and the ''cool'' temperature field decreases moving to the cool origin [Figs. 5(d) and 5(e)]. When the growth front approaches the crystal side edge, the ''cool'' temperature field begins the back motion to the surface [Fig. 5(f)]. This is the full cycle of the epitaxial compact growth.

For this cycle growth process, the quasistationary limit $\partial T/\partial t \to 0$ and Eq. (14) are unacceptable. Therefore, it is hardly possible to obtain an analytic criterion for the compact-DBM transition. To find a solution, we applied the method of the dimension analysis to this problem. We considered the following approximation for function $\eta_2(\xi,\lambda)$:

$$\eta_2(\xi,\lambda) \sim \xi^\alpha (1-\xi)^\beta (1+\lambda)^\gamma, \qquad (19)$$

where $\alpha$, $\beta$, and $\gamma$ are unknown parameters that obey the condition $\alpha+\beta+\gamma=0$. The values of parameters $\alpha=\gamma=\frac{1}{2}$, $\beta=-1$ were determined from MC simulations. As a result, we obtained the following formula:

$$\eta_2(\xi,\lambda) \sim \frac{\sqrt{(T_A+L/C-T_O)T_{\text{surf}}}}{T_A-T_{\text{surf}}-T_O} \sim \frac{\sqrt{\xi(1+\lambda)}}{1-\xi}. \qquad (20)$$

### C. Morphological diagram

Equations (17), (20) give the complete information about the morphology of the growing aggregate. In general case the morphological type is a function of three variables: $\xi$, $\lambda$, and $\eta$. However, in most cases the parameter $\lambda < 1$ so its functional dependence in Eqs. (17), (20) is rather weak in comparison to the dependence of the parameter $\xi$ which is crucial for the pattern type. Thus, it is quite acceptable to illustrate the morphology diagram in two coordinates: $\xi$ and $\eta$. The 2D restriction ($\lambda =$ const) of the diagram is presented in Fig. 6. The figure shows the three kinds of numerically obtained growth patterns. The fractal structures are observed at $\eta > \eta_1(\xi,\lambda)$, the compact growth occurs at $\eta < \eta_2(\xi,\lambda)$, and the intermediate case $\eta_2(\xi,\lambda) < \eta < \eta_1(\xi,\lambda)$ corresponds to the DBM patterns. The diagram sections ($\xi =$ const and $\eta =$ const) demonstrate the observed morphological evolutions [Figs. 2 and 3].

## ACKNOWLEDGMENT

We would like to thank Mr. George Olson for useful discussions and helpful comments.